\documentclass[conference]{IEEEtran}
\IEEEoverridecommandlockouts
\usepackage{cite}
\usepackage{amsmath,amssymb,amsfonts}
\usepackage{algorithmic}
\usepackage{graphicx}
\usepackage{textcomp}
\usepackage{xcolor}
\usepackage{booktabs} 
\usepackage{listings}
\usepackage{soul}
\usepackage{xspace}
\usepackage{graphicx}
\usepackage{ifthen}
\usepackage[normalem]{ulem} 
\usepackage{xcolor,colortbl}
\usepackage{hyperref}

\newboolean{showedits}
\setboolean{showedits}{true} 
\ifthenelse{\boolean{showedits}}
{
	\newcommand{\del}[1]{\textcolor{red}{\sout{#1}}} 
	\newcommand{\nbe}[3]{
		{\colorbox{#3}{\bfseries\sffamily\scriptsize\textcolor{white}{#1}}}
		{\textcolor{#3}{\sf\small$\blacktriangleright$\textit{#2}$\blacktriangleleft$}}}
}{
	\newcommand{\del}[1]{} 
	
	\newcommand{\nbe}[3]{}
}

\newboolean{showcomments}
\setboolean{showcomments}{true} 
\newcommand{\id}[1]{$-$Id: scgPaper.tex 32478 2010-04-29 09:11:32Z oscar $-$}

\ifthenelse{\boolean{showcomments}}
 {
 	\newcommand{\nbc}[3]{
 		{\colorbox{#3}{\bfseries\sffamily\scriptsize\textcolor{white}{#1}}}
		{\textcolor{#3}{\sf\small$\blacktriangleright$\textit{#2}$\blacktriangleleft$}}}
	
 }{
 	\newcommand{\nbc}[3]{}
 	
 }

\usepackage[most]{tcolorbox}
\ifthenelse{\boolean{showedits}}
{
  \newtcolorbox{inserted}{%
       title=Inserted text:,
       colframe=blue,colback=blue!5!white,
       breakable,
       leftrule=0mm, 
       bottomrule=0mm,
       rightrule=0mm,
       toprule=0mm,
       arc=0mm, outer arc=0mm,
       oversize
  }
  \newtcolorbox{deleted}{%
       title=Deleted text:,
       colframe=red,colback=red!5!white,
       breakable,
       leftrule=0mm, 
       bottomrule=0mm,
       rightrule=0mm,
       toprule=0mm,
       arc=0mm, outer arc=0mm,
       oversize
  }
  \newtcolorbox{refactored}{%
       title=Rewritten text:,
       colframe=blue,colback=red!5!white,
       breakable,
       leftrule=0mm, 
       bottomrule=0mm,
       rightrule=0mm,
       toprule=0mm,
       arc=0mm, outer arc=0mm,
       oversize
  }
}{

}

\newboolean{hide}
\setboolean{hide}{true} 
\ifthenelse{\boolean{hide}}
{
 \newcommand{\hide}[1]{}
}{
 \newcommand\hide[1]{\nbc{HIDE}{#1}{red}}
}

\newcommand{\commented}[1]{}

\newcommand{\eg}{\emph{e.g.,}\xspace}
\newcommand{\ie}{\emph{i.e.,}\xspace}
\newcommand{\etal}{\emph{et al.}\xspace}

\newboolean{isblinded}
\setboolean{isblinded}{true}
\ifthenelse{\boolean{isblinded}}
{\newcommand\blind[1]{BLINDED\xspace}}
{\newcommand\blind[1]{#1\xspace}}

\definecolor{source}{gray}{0.95}

\usepackage[T1]{fontenc} 

\lstnewenvironment{codesnippet}{%
	\lstset{%
		frame=single,
		framerule=0pt,
		mathescape=false
	}
}{}

\newcommand{\CG}{CogniCrypt\xspace}
\newcommand{\GH}{GitHub\xspace}
\newcommand{\CE}{CryptoExplorer\xspace}

\begin{document}

\title{\CE: An Interactive Web Platform Supporting Secure Use of Cryptography APIs}

\author{\IEEEauthorblockN{Mohammadreza Hazhirpasand}
\IEEEauthorblockA{SCG, University of Bern\\
Bern, Switzerland\\
mohammadreza.hazhirpasand@inf.unibe.ch}
\and
\IEEEauthorblockN{Mohammad Ghafari}
\IEEEauthorblockA{SCG, University of Bern\\
Bern, Switzerland\\
mohammad.ghafari@inf.unibe.ch}
\and
\IEEEauthorblockN{Oscar Nierstrasz}
\IEEEauthorblockA{SCG, University of Bern\\
Bern, Switzerland\\
oscar.nierstrasz@inf.unibe.ch}
}

\IEEEoverridecommandlockouts
\IEEEpubid{\makebox[\columnwidth]{\textbf{Preprint -- SANER 2020}\hfill} \hspace{\columnsep}\makebox[\columnwidth]{ }}

\makeatletter                  
\def\mdseries@tt{m}      
\makeatother                   

\maketitle

\begin{abstract}

Research has shown that cryptographic APIs are hard to use.
Consequently, developers resort to using code examples available in online information sources that are often not secure.

We have developed a web platform, named \CE, stocked with numerous real-world secure and insecure examples that developers can explore to learn how to use cryptographic APIs properly.
This platform currently provides 3\,263 secure uses, and 5\,897 insecure uses of Java Cryptography Architecture mined from 2\,324 Java projects on \GH.

A preliminary study shows that \CE provides developers with secure crypto API use examples instantly, developers can save time compared to searching on the internet for such examples, and they learn to avoid using certain algorithms in APIs by studying misused API examples.

We have a pipeline to regularly mine more projects, and, on request, we offer our dataset to researchers.

\end{abstract}

\begin{IEEEkeywords}
Cryptography, security, code analysis
\end{IEEEkeywords}

\section{Introduction}
\label{sec:intro}
Employing cryptographic APIs (or ``crypto APIs'') correctly is a troublesome task for developers.
For instance, a recent study of Java Cryptography Architecture (JCA) in 2\,324 projects on GitHub showed that more than 72\% of the projects suffer from at least one cryptographic misuse, and of 1\,578 distinct developers who committed cryptography code, 41\% have always misused cryptography~\cite{hazhirpasand2019impact}.

The widespread misuse of crypto APIs has manifold reasons.
API documentation, as the official source of learning how to use crypto APIs, lacks security-related hints~\cite{Yasemin2017}.
Code snippets obtained from online information sources are untrustworthy too.
A study of 1.3 million Android apps showed that 196\,403 (\ie15\%) used vulnerable code snippets that were very likely copied from the Stack Overflow website\cite{Fischer2017}.
Examination of 217\,818 Stack Overflow posts also showed that 31\% suffer from potential API misuses that could lead to unexpected behavior such as program crashes and resource leaks \cite{zhang2018}.
There exist several tools to assist developers in using crypto APIs correctly~\cite{crypgaurd,Egele}.
These tools nevertheless need to be installed, work in a specific programming environment, and each is complex to learn.
What's worse, not all developers are aware of such tools.

%
We introduce \CE,\footnote{\url{https://www.crypto-explorer.com}} a web platform that developers can use to reliably explore secure crypto code examples mined from open-source projects, and compare them against insecure code uses.
Developers can either look for a particular API use or explore examples that are similar to a given crypto code snippet.
This platform mines open-source projects to increase the number of API usage examples, and opens a potential bug report in each project that suffers from crypto misuses.
At the time of writing, \CE provides 3\,263 secure uses, and 5\,897 insecure uses of Java Cryptography Architecture (JCA) mined from 2\,324 Java projects on \GH.
A preliminary user study shows that this platform helps developers to find secure crypto examples, and learn how to properly use crypto APIs.

We believe the advantage of using \CE is twofold.
Not only will the platform help developers to find secure examples of crypto APIs or learn from insecure examples, but they also do not need to search for an appropriate analysis tool and struggle with installation issues or spend time on the internet to look for secure examples.
Furthermore, as the \CE pipeline gathers more crypto usages behind the scene at regular intervals, the dataset, available on request, will be useful for researchers who want to conduct any large-scale analysis in this domain.

The remainder of this paper is structured as follows.
In \autoref{sec:motive}, we explain various types of mistakes in using crypto APIs.
In \autoref{sec:ce} we introduce the \CE platform and its current use cases.
In \autoref{sec:study} we present a preliminary user study,
and discuss related work in \autoref{sec:related}.
We conclude this paper in \autoref{sec:conclusion}.
\section{Motivation}
\label{sec:motive}

The method in \autoref{lst:codesnippet} presents a password-based encryption.
Lines 2 to 7 derive the cryptographic key \texttt{cipherkey} from a password \texttt{pwd}.
The remaining three lines perform the encryption of \texttt{plaintext} using the key.
The encryption is, however, insufficiently secure due to several mistakes of different kinds in the key derivation.

\newpage

\emph{Wrong Type}: In Java, passwords should not be stored as String objects as they are immutable and therefore cannot be overwritten once they are no longer used.
That is why the constructor call of \texttt{PBEKeySpec} requires the password to be passed as a character array.

\emph{Wrong object}:
The second parameter of the \texttt{PBEKeySpec} constructor is insecure as it must be random to fulfill its purpose as a salt, but it has been defined as a constant array one line above.
According to this misuse, a crypto object that flows into the member method of another does not fulfill a certain expected requirement.

\begin{lstlisting}[language=Java, caption=Various types of crypto API misuses, label=lst:codesnippet]
public byte [ ] encrypt(byte [ ] plaintext, String pwd) {
	byte [ ] salt = {15, -12, 94, 0, 12, 3, -65, 73,-1, -84, -35};
	PBEKeySpec spec = new PBEKeySpec (pwd.toCharArray(), salt, 100);

	SecretKeyFactory skf = SecretKeyFactory.getInstance("PBKDF2WithHmacSHA256");
	byte [ ] keyMaterial = skf.generateSecret(spec).getEncoded();
	SecretKeySpec cipherKey = new SecretKeySpec(keyMaterial, "AES");

	Cipher ciph = Cipher.getInstance("AES/ CBC/ PKCS5Padding");
	ciph.init(Cipher.ENCRYPT_MODE, cipherKey);
	return ciph.doFinal(plaintext);
} 
\end{lstlisting}

\emph{Wrong constraint}: The third parameter of \texttt{PBEKeySpec} is also not secure.
It specifies the number of iterations that the password is hashed in order to derive a cryptographic key.
This iteration count parameter however should not be lower than 1\,000 while in this example it is only 100 \cite{turan2010recommendation}.
This type of violation concerns wrong values for integer or string objects, like key sizes, algorithm names, or iteration counts.

\emph{Forbidden call}: 
A method may be forbidden and should not be called if it is outdated but left in the API for legacy reasons.
One such forbidden method is the \texttt{PBEKeySpec} constructor call in this snippet.
The call is discouraged because it does not set the key size of the key derived from its password.

\emph{Incomplete operation}: A developer should call \texttt{PBEKeySpec.clearPassword()}  after the key has been created.
When this method is executed, the password within the object is nullified.
This code snippet, however, fails to make this call, leading to an incomplete operation in finalizing the whole path for the desired cryptographic purpose.
This misuse occurs when developers forget to call all required methods on an object.

\emph{Incomplete order}: This misuse is signaled when a required sequence of method calls on a crypto object is not respected.
This would be the case if the call to \texttt{ciph.init()} were to be omitted.

Such categories of errors commonly occur in using cryptographic APIs. 
We analyze a significant number of Java applications in order to present a large dataset of analyzed projects to the research community and help developers to use such APIs correctly.

\section{\CE}
\label{sec:ce}

In this section we present the \CE web platform that developers can use to explore real-world crypto API uses mined from open-source projects.
Developers can search crypto APIs, explore secure and insecure API uses, and compare them.

\begin{figure}
\centering
\centerline{\includegraphics[width=1\linewidth]{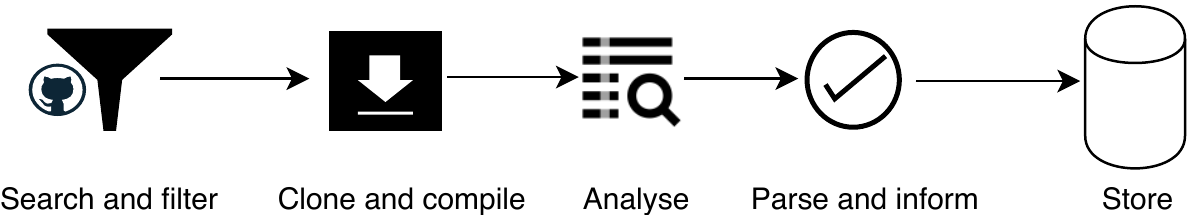}}
\caption{The workflow of \CE
}\label{fig:cryptoexplorer1}
\end{figure}

We explain the workflow supported by the tool as presented in \autoref{fig:cryptoexplorer1}, and explain the current use cases.

\subsection{Workflow}
\CE automatically grows its database of cryptographic examples by finding cryptographic-related projects and analyzing and storing the cryptographic API examples.
The pipeline consists of five major steps to add one cryptographic example.
We use the cron scheduling daemon to automatically schedule and execute bash scripts.

\subsubsection{Search and filter}
\hide{To add more cryptographic API usages, we look for projects in public repositories whose code employs cryptographic APIs.
To this end, we currently analyze open source projects hosted on \GH as it offers an APIs to facilitate the process of finding cryptographic-related projects.}
To add more cryptographic API usages, we look for current open source projects hosted on \GH.
First, we look for Java projects using \GH's repository search API.
We exclude forked projects to avoid cloning duplicated projects.
We then use \GH's code search API to search for JCA APIs inside Java projects.
Finally, we store the addresses of Java projects whose code contains cryptographic APIs.

\subsubsection{Clone and compile}
We clone and compile projects to perform static analysis.
We clone identified Java projects containing JCA APIs.
To build the projects, we check for the presence of the Project Object Model (POM) file in the project's path, and if it exists, we proceed with compilation.
The POM file is an eXtensible Markup Language (XML) representation of a Maven project.
We use the Maven build tool for the compilation process and we skip projects in which dependencies cannot be resolved.
Lastly, we neither download the forked version of a project nor a project twice for analysis.

\subsubsection{Analyse}
We currently employ \CG, a static-analysis tool tailored to find a wide range of misuses of JCA APIs~\cite{kruger2017cognicrypt}.
It takes a target program and specification rules (\eg  method-call patterns, parameter constraints and secure compositions of cryptography-related classes) as input, and evaluates the program's correctness with respect to these rules.
The tool returns secure and buggy API usages.
We specified a time period, \ie 15 minutes, for \CG to run the analysis to limit time and resources used on the server.


In future, we plan to add more analysis tools and present their results in \CE.

\subsubsection{Parse and inform}
We extract information from the analysis report to present to developers via \CE.
In particular, for every project, we extract which cryptographic API was (mis)used, the reason for being misused, the corresponding file name, and the line number of each detected (mis)use.
With the help of the \textit{git blame} command we also identify the last developer who committed the code associated with each API use, as well as the commit time.

\newpage

Afterwards, we create issues on the \GH page's of projects to report the potential misuses. 
In case the project owner decided to disable issues for the project, we send an email to the developers who committed the API misuse.
Each issue report includes help instructions related to the type of misuse, line number, and file path.

In order to curate a healthy dataset, we check responses to each issue and, in case of a false positive, we adapt the entry in \CE's dataset.
As we need to manually analyze the responses, we mine 100 projects weekly at the moment.
At the moment, we do not re-examine the repositories automatically once we notify them concerning the crypto misuses. 

\subsubsection{Store}
We store the analysis results in a database, tracking elements such as the filename, the crypto API name, the API call line number, the user-defined function where JCA APIs were used, and whether the API use is secure or not.
\CE is designed to support multiple languages: it must be configured with the API host language, build tools, and static analysis tools.
If a specific crypto algorithm is found to be vulnerable (\eg \texttt{SHA-256}), 
it is feasible to mark the specific crypto algorithm's usages as buggy in the database.

\autoref{tab:projects} presents the current numbers of secure and buggy projects and commits, and their totals.
There is an average of 1.7 distinct API usages per project with a standard deviation of 1.3 and an average number of 3.9 commits per project with a standard deviation of 7.4.

\begin{table}[h]
\center
\caption {The status of projects and commits} \label{tab:projects} 
\begin{tabular}{@{}lrrr@{}}
  & Secure & Buggy & Total \\ \midrule
Projects & 642   & 1,682  & 2,324  \\
Commits (LoC)  & 3,263  & 5,897  & 9,160  \\ 
\end{tabular}
\end{table}

\subsection{Usage Scenario}

\begin{figure*}
\centering
\centerline{\includegraphics[width=0.95\linewidth]{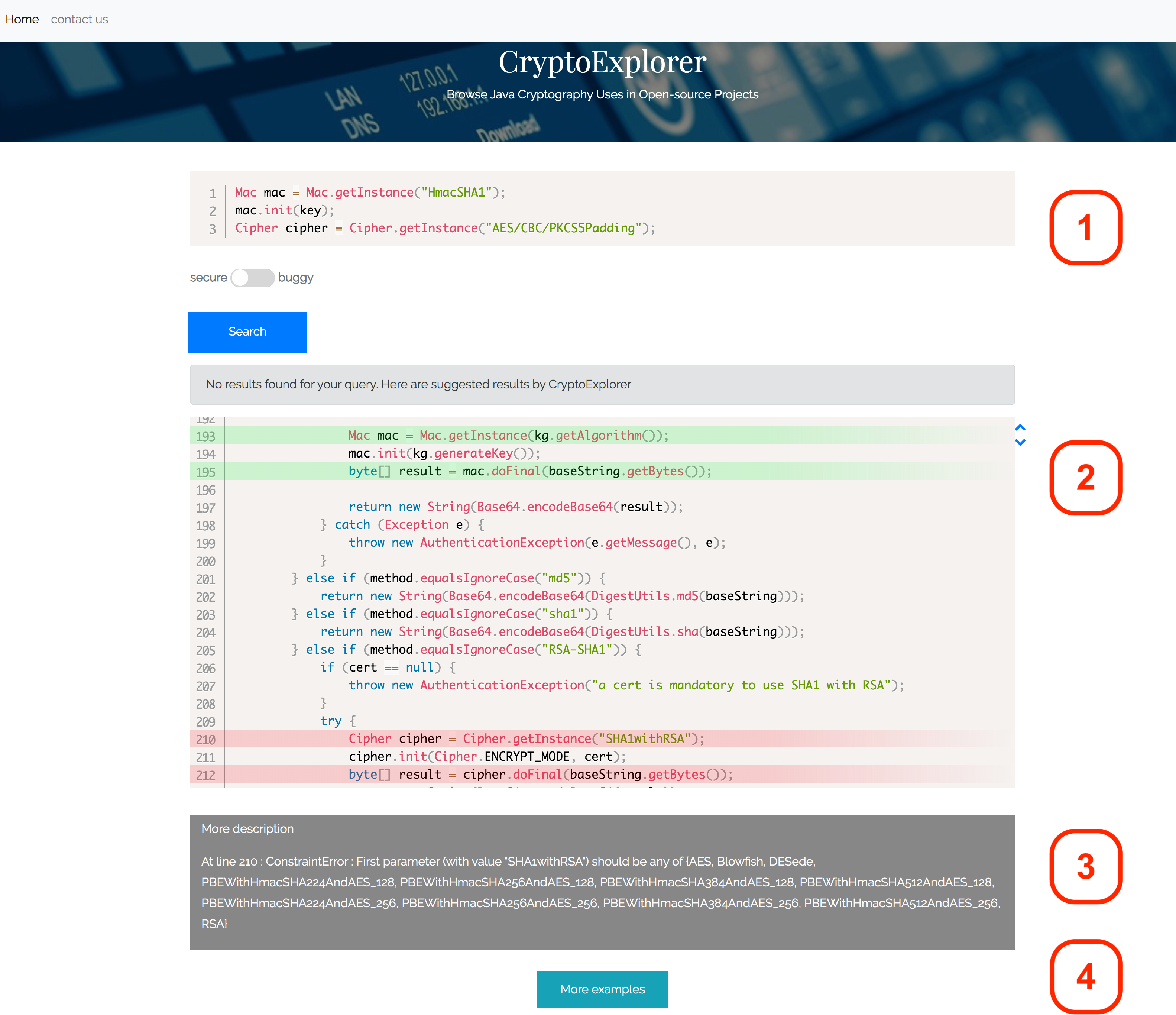}}
\caption{Exploring code examples based on a given code snippet
}\label{fig:cryptoexplorer2}
\end{figure*}

The user interface of \CE, shown in \autoref{fig:cryptoexplorer2}, is simple to use, and just a few options need to be adjusted to tune the search query. 
Developers can either directly visit the publicly available website of the \CE, or use an Eclipse plugin that we have developed to interact with \CE from within the IDE.
The simplicity of the plugin only requires developers to select either an entire Java file or part of one and click on the plugin's icon to open \CE in a web browser.

\CE allows developers to search for example usages of a particular cryptographic API.
For instance, a user might input the \texttt{MessageDigest} API name to see how this API is (mis)used in different examples in order to circumvent the lack of usage examples and security hints in the official Java documentation. 
Developers may also input a piece of cryptography code that they would like to evaluate, or to use as a query to find similar crypto usages.
They can choose to explore only secure uses or buggy uses. 
In case \CE does not find any example, it suggests examples where both secure and buggy uses of the queried APIs exist, distinguishable by green and red colors, respectively.

When developers input sample code to be used to search for usages of cryptographic APIs, \CE identifies the cryptographic APIs in the code, and presents developers with code examples that use the same APIs.
For better readability, we rank those files higher where crypto API usages are close to each other in a file and are in the same user-defined method.
We use standard deviation to compute how close API usages are in a file.
As each API could be misused in a few ways, \CE does not return hundreds of examples to users with several identical key messages.
In case users are interested to study more APIs examples, they can navigate more examples.

For instance, in \autoref{fig:cryptoexplorer2}, a user has entered a piece of code that aims to conduct file encryption.
\CE recognizes the \texttt{Cipher} and \texttt{Mac} cryptographic APIs in the code (1).
Once the user hits the ``Search'' button,  \CE returns code examples that simultaneously use the same APIs listed in the search code (2).
In this example, \CE could not find any secure example where both APIs were used securely, and it suggested an example where the APIs were used in both secure and buggy ways.
The misused API is highlighted by a red linear gradient color and the secure API usage is highlighted by a green linear gradient color. 
In the case of navigating large files, there are two buttons that assist users to jump to the highlighted lines quickly. 
Under each returned example, users can read related information regarding the example's misuses (3).
Finally, users can navigate for more examples (4).

\section{User Experience}
\label{sec:study}
We investigated the experience of users interacting with \CE to understand to which extent it supports developers to properly use crypto APIs.

\subsection{Methodology}

We conducted semi-structured interviews with four participants who used this platform.
They willingly chose to participate without being paid and
all had an academic background in computer science (\ie two bachelors, and two Ph.D. students).
The participants had at least 2 years of experience in Java programming. 
They all used \texttt{MessageDigest} to produce hashes, and they were familiar with cryptography concepts. 
However, it was not their daily job to write cryptographic-related code in Java or any other languages.

We presented \CE to the participants, and explained its features. Then we asked them to accomplish the following two tasks using \CE, while they could consult official JCA documentation as well:

\begin{itemize}

\item \textbf{Task1.} 
Tell us of two security concerns that one should consider when using the \texttt{MessageDigest} API to generate a hash.
%

\item \textbf{Task2.} 
Find security issues in a given crypto code snippet that uses the \texttt{Cipher} API to conduct file encryption, and explain how to resolve them.
This task first must be done by using any resources on the internet and then with the help of  \CE.

\end{itemize}

We asked participants to think aloud while working on each task.
In the end, we interviewed them regarding the difficulties that they experienced.

\subsection{Results}

To accomplish the first task, participants had to search the \texttt{MessageDigest} API, and  explore 20 similar code examples. Every participant succeeded to complete this task, on average within seven minutes.
They stated that they had to read all of the returned examples as the API was used in a different ways (\eg initialized with different hashing algorithms).
A participant suggested adding an option to decide whether to exclude an example due to false positives. 
Such examples may present different scenarios that are not cryptography-related or tool's mistake.
When we asked whether they know why a particular security issue exists, they stated that sometimes this was not directly evident from a misuse itself. 
They had to read the information below each example, and in a few cases, they stated the information is not sufficiently expressive.
For instance, one participant did not know why using \texttt{SHA-1} is not secure.
One participant said that providing external links for each misuse could help demystify the reason behind each misuse.
A participant explored more examples by clicking on the more examples button, and we realized that he could not completely benefit from examples that had variables whose definitions were missing in the provided Java file. 
He suggested excluding examples whose information spans more than one Java file and providing a button to report such examples.
All in all, all participants figured out what hashing algorithms, \eg \texttt{MD5} or \texttt{SHA1}, can be problematic or the importance of calling the \texttt{Digest} and \texttt{Update} methods in \texttt{MessageDigest}.

To accomplish the second task, participants first used the internet for 20 minutes to find out the misuses in the code snippet.
Only one participant could realize where the problem lies. 
Then, they used \CE and completed the task on average in eight minutes.
They all had to go through six similar examples to learn the correct and wrong way of using the API.
The found that mainly because the algorithm/mode/padding string in the \texttt{getInstance} method of the \texttt{Cipher} API accepts several algorithms, mode, and padding modes. 
Participants stated that the buttons for jumping to affected lines help immensely as sometimes files contain hundreds of lines of codes. They also suggested that a search feature for each example would ease the problem of finding variables.
We also discovered that different types of misuses have different levels of difficulty for users to understand.
For instance, the incomplete operation error type required developers to carefully read a misuse example, while a constraint error was often clear to participants only after looking at a misuse example.

Finally, we asked participants about their experience with \CE, and how much easier it is compared to searching for misuses on the internet.
They all agreed that \CE facilitates learning how a crypto API should be used correctly by providing real-world analyzed examples, highlighting the lines, and presenting information related to each example.
They also pointed out that it is extremely hard to find topics on Stack Overflow or a particular website regarding misuses of a specific cryptographic API and that they cannot trust the provided cryptographic code snippets.

\section{Related Work}
\label{sec:related}

CryptoLint, developed by Egele \etal, checks real-world Android applications for the violation of six security rules \cite{Egele}.
They succeeded to find 10\,327 of 11\,748 Android applications analyzed by CryptoLint that use cryptographic APIs exposed to at least one mistake.
CryptoLint is not yet open source.

Yong Li \etal proposed iCryptoTracer, which performs a combination of static and dynamic analysis on iOS applications \cite{iosstatic}.
Their research showed that nearly 65.3\% of the examined applications suffered from a cryptographic misuse.

Rahaman \etal describe CRYPTOGUARD, a deployment-quality static analysis tool to identify Java cryptographic misuses \cite{crypgaurd}.
They provide contextual refinements for false positive reduction, on-demand flow-sensitive, and context-sensitive analysis.

Kim \etal introduced a code search engine that merges results from  API documents with 
code example summaries, mined from the web \cite{kim2010towards}.
However, it is not tailored to mine secure examples.

Kr\"uger \etal presented a tool called \CG, an Eclipse plugin that empowers developers to identify cryptographic misuses in Java code\cite{kruger2017cognicrypt}.




In summary, \CE differs from existing work in several ways: (1) it maintains secure crypto API usage examples that practitioners can use to learn how to  properly use crypto APIs, and researchers can use to benchmark analysis tools in this domain; (2) it continuously mines projects on \GH, and can notify project developers of existing cryptographic misuses. 


\section{Conclusion}
\label{sec:conclusion}

We have presented \CE, a web platform to search for real-world crypto API (mis)uses in open-source projects.
It currently provides hundreds of code examples mined from 2\,324 Java projects on \GH.
A preliminary study showed that \CE helps developers to find secure crypto examples, and learn how to properly use crypto APIs by examining examples of correct uses and misuses.
Moreover, as the dataset of \CE is growing, it can be useful for researchers to conduct other related studies.

We plan to provide more useful explanations with respect to each API misuse, to further help developers to comprehend the reason underpinning each problem.
We are considering adding a feature for users to compare secure and buggy examples side-by-side.
We intend to provide the possibility of exploring code examples based on crypto scenarios.
Finally, we will expand \CE with both more programming languages and other crypto libraries.


\section*{Acknowledgment}
We gratefully acknowledge the financial support of the 
Swiss National Science Foundation for the project ``Agile Software Assistance'' (SNSF project No.\,200020-181973, Feb.\,1, 2019 -- April 30, 2022).
We also thank CHOOSE, the Swiss Group for Original and Outside-the-box Software Engineering
of the Swiss Informatics Society, for its financial contribution to the presentation of this paper.


\bibliographystyle{IEEEtran}
\bibliography{cryptoexplorer}

\begin{thebibliography}{10}
\providecommand{\url}[1]{#1}
\csname url@samestyle\endcsname
\providecommand{\newblock}{\relax}
\providecommand{\bibinfo}[2]{#2}
\providecommand{\BIBentrySTDinterwordspacing}{\spaceskip=0pt\relax}
\providecommand{\BIBentryALTinterwordstretchfactor}{4}
\providecommand{\BIBentryALTinterwordspacing}{\spaceskip=\fontdimen2\font plus
\BIBentryALTinterwordstretchfactor\fontdimen3\font minus
  \fontdimen4\font\relax}
\providecommand{\BIBforeignlanguage}[2]{{%
\expandafter\ifx\csname l@#1\endcsname\relax
\typeout{** WARNING: IEEEtran.bst: No hyphenation pattern has been}%
\typeout{** loaded for the language `#1'. Using the pattern for}%
\typeout{** the default language instead.}%
\else
\language=\csname l@#1\endcsname
\fi
#2}}
\providecommand{\BIBdecl}{\relax}
\BIBdecl

\bibitem{hazhirpasand2019impact}
M.~Hazhirpasand, M.~Ghafari, S.~Kr{\"u}ger, E.~Bodden, and O.~Nierstrasz, ``The
  impact of developer experience in using {Java} cryptography,'' in \emph{2019
  ACM/IEEE International Symposium on Empirical Software Engineering and
  Measurement (ESEM)}.\hskip 1em plus 0.5em minus 0.4em\relax IEEE, 2019, pp.
  1--6.

\bibitem{Yasemin2017}
Y.~{Acar}, M.~{Backes}, S.~{Fahl}, S.~{Garfinkel}, D.~{Kim}, M.~L. {Mazurek},
  and C.~{Stransky}, ``Comparing the usability of cryptographic {APIs},'' in
  \emph{2017 IEEE Symposium on Security and Privacy (SP)}, May 2017, pp.
  154--171.

\bibitem{Fischer2017}
F.~{Fischer}, K.~{B\"ottinger}, H.~{Xiao}, C.~{Stransky}, Y.~{Acar},
  M.~{Backes}, and S.~{Fahl}, ``Stack overflow considered harmful? the impact
  of copy paste on {Android} application security,'' in \emph{2017 IEEE
  Symposium on Security and Privacy (SP)}, May 2017, pp. 121--136.

\bibitem{zhang2018}
T.~{Zhang}, G.~{Upadhyaya}, A.~{Reinhardt}, H.~{Rajan}, and M.~{Kim}, ``Are
  code examples on an online {QA} forum reliable?: A study of {API} misuse on
  {Stack Overflow},'' in \emph{2018 IEEE/ACM 40th International Conference on
  Software Engineering (ICSE)}, May 2018, pp. 886--896.

\bibitem{crypgaurd}
S.~Rahaman, Y.~Xiao, K.~Tian, F.~Shaon, M.~Kantarcioglu, and D.~Yao,
  ``{CHIRON:} deployment-quality detection of {Java} cryptographic
  vulnerabilities,'' \emph{CoRR}, vol. abs/1806.06881, 2018.

\bibitem{Egele}
M.~Egele, D.~Brumley, Y.~Fratantonio, and C.~Kruegel, ``An empirical study of
  cryptographic misuse in {Android} applications,'' in \emph{Proceedings of the
  2013 ACM SIGSAC Conference on Computer Communications Security}, ser. CCS
  '13.\hskip 1em plus 0.5em minus 0.4em\relax New York, NY, USA: ACM, 2013, pp.
  73--84.

\bibitem{turan2010recommendation}
M.~S. Turan, E.~Barker, W.~Burr, and L.~Chen, ``Recommendation for
  password-based key derivation,'' \emph{NIST special publication}, vol. 800,
  p. 132, 2010.

\bibitem{kruger2017cognicrypt}
S.~Kr{\"u}ger, S.~Nadi, M.~Reif, K.~Ali, M.~Mezini, E.~Bodden, F.~G{\"o}pfert,
  F.~G{\"u}nther, C.~Weinert, D.~Demmler \emph{et~al.}, ``Cognicrypt:
  Supporting developers in using cryptography,'' in \emph{Proceedings of the
  32nd IEEE/ACM International Conference on Automated Software
  Engineering}.\hskip 1em plus 0.5em minus 0.4em\relax IEEE Press, 2017, pp.
  931--936.

\bibitem{iosstatic}
Y.~Li, Y.~Zhang, J.~Li, and D.~Gu, ``{iCryptoTracer}: Dynamic analysis on
  misuse of cryptography functions in {iOS} applications,'' in \emph{Network
  and System Security}, M.~H. Au, B.~Carminati, and C.-C.~J. Kuo, Eds.\hskip
  1em plus 0.5em minus 0.4em\relax Cham: Springer International Publishing,
  2014, pp. 349--362.

\bibitem{kim2010towards}
J.~Kim, S.~Lee, S.-w. Hwang, and S.~Kim, ``Towards an intelligent code search
  engine,'' in \emph{Twenty-Fourth AAAI Conference on Artificial Intelligence},
  2010.

\end{thebibliography}

\end{document}